\documentclass[]{aastex631}
\usepackage{physics}
\usepackage{multirow}

\shorttitle{SWASTi-SW}
\shortauthors{Mayank et al.}

\begin{document}

\title{SWASTi-SW: Space Weather Adaptive SimulaTion framework for Solar Wind and its relevance to ADITYA-L1 mission}

\correspondingauthor{Prateek Mayank}
\email{phd2001121004@iiti.ac.in}

\author[0000-0001-8265-6254]{Prateek Mayank}
\affiliation{Department of Astronomy, Astrophysics and Space Engineering,\\ Indian Institute of Technology Indore,\\ Khandwa Road, Simrol, 453552, India}

\author[0000-0001-5424-0059]{Bhargav Vaidya}
\affiliation{Department of Astronomy, Astrophysics and Space Engineering,\\ Indian Institute of Technology Indore,\\ Khandwa Road, Simrol, 453552, India}

\author[0000-0003-2693-5325]{D. Chakrabarty}
\affiliation{Space and Atmospheric Sciences Division,\\ Physical Research Laboratory,\\ Ahmedabad, 380009, India}



\begin{abstract}

Solar wind streams, acting as background, govern the propagation of space weather drivers in the heliosphere, which induce geomagnetic storm activities. Therefore, predictions of the solar wind parameters are the core of space weather forecasts. This work presents an indigenous three-dimensional (3D) Solar Wind model (SWASTi-SW). This numerical framework for forecasting the ambient solar wind is based on a well-established scheme that uses a semi-empirical coronal model and a physics-based inner heliospheric model. This study demonstrates a more generalized version of Wang-Sheeley-Arge (WSA) relation, which provides a speed profile input to the heliospheric domain. Line-of-sight observations of GONG and HMI magnetograms are used as inputs for the coronal model, which in turn, provides the solar wind plasma properties at 0.1 AU. These results are then used as an initial boundary condition for the magnetohydrodynamics (MHD) model of the inner heliosphere to compute the solar wind properties up to 2.1 AU. Along with the validation run for multiple Carrington rotations, the effect of variation of specific heat ratio and study of stream interaction region (SIR) is also presented. This work showcases the multi-directional features of SIRs and provides synthetic measurements for potential observations from the Solar Wind Ion Spectrometer (SWIS) subsystem of Aditya Solar wind Particle EXperiment (ASPEX) payload on-board ISRO's upcoming solar mission Aditya-L1.

\end{abstract}




\section{Introduction} \label{sec:intro}

The study of influence of the Sun on our Earth has become an essential area of research globally, known as Space Weather. The effects of space weather can disrupt electric power supply, perturb navigation systems, interrupt satellite functionality, and can be hazardous to astronaut health. The extremely energetic space weather events could even impact the global economy. Therefore, the importance of space weather forecasting is recognised not only by the research community but also by government and industry stakeholders \citep{Schrijver2015}.
\par
To mitigate the adverse effects of space weather, it is crucial to understand the underlying physics to prepare our timely rational response. The existing observatories provide the details of near solar surface region (e.g., SoHO, STEREO, SDO, GONG, etc.), near-Earth region (e.g., ACE, WIND, DSCOVR, etc.) and the inner heliospheric region (e.g., Parker Solar Probe, Solar Orbiter, etc.). These observatories present a good starting point for the study but don't produce the required insight of arrival time of the energetic space weather events. So, to bridge this gap, numerical models are necessary, which could use the observed data and forecast hazardous events like coronal mass ejections (CMEs), solar energetic particles (SEPs), stream interaction regions (SIRs), etc. Solar Wind streams, acting as a background, govern the propagation of these events in the heliosphere and drive geomagnetic storm activities. Therefore, predictions of the solar wind parameters are the core of space weather forecasts.
\par
The method of inner-heliospheric modeling of solar wind can be broadly classified into three categories: empirical, semi-empirical and MHD based simulation. Empirical models (e.g., PDF \citep{Bussy-Virat2014}, PROJECTZED \citep{Riley2017}, AnEn \citep{Owens2017}, etc.) use a probabilistic forecasting approach, which is framed by analyzing solar wind observations at Sun-Earth L1 Lagrangian point. The semi-{empirical} models (e.g., ESWF \citep{Reiss2016}, WSA \citep{Arge2000}, etc.) employ an empirical relation of solar wind speed based on the observation of coronal holes. On the other hand, simulation models (e.g., MAS \citep{Riley2001}, ENLIL \citep{Odstrcil2003}, SWMF \citep{Toth(2005)}, SWIM \citep{Feng2010}, SUSANOO \citep{Shiota2014}, EUHFORIA \citep{Pomoell2018}, \citet{Narechania2021} etc.) are physics-based models which use photospheric magnetograms to determine plasma properties in the heliosphere. According to the assessment of \citet{MacNeice2018}, the accuracy of empirical models surpasses the results of semi-empirical and simulation models. The empirical models are relatively less intensive computationally and give more accurate results. However, these models offer a limited number of plasma properties that too only at the L1 location. Though simulation models are computationally expensive, these models provide an understanding of fundamental physics which is essential for preparing a sensible response against the space weather effects.
\par
A coupled two domain procedure is followed by almost all simulation models. In some cases MHD codes are used for both domains (e.g., MAS, SWMF, ENLIL-MHD, SWIM, etc.) and in some cases, WSA model is used in the coronal domain (e.g., ENLIL-WSA, SUSANOO, EUHFORIA, \citet{Narechania2021}).
\par
This work is the first step towards our central objective, i.e., to develop a full-fledged data-driven \textit{Space Weather Adaptive SimulaTion} (SWASTi) framework. In this paper, we present a solar wind model, SWASTi-SW, which is the first part of this modular framework. This newly developed physics-based solar wind model uses an updated WSA approach in coronal domain and MHD code in inner-heliospheric domain. The MHD domain uses {\sc pluto} code \citep{Mignone2007} to compute the plasma properties in the inner-heliosphere. An earlier assessment of usage of {\sc pluto} code for solar wind prediction was done in two dimensional pilot study by \citet{Kumar2020}, in which they compared the results with other extrapolation models using WSA relation. In this three dimensional work, a more generalized version of WSA relation is used along with more robust and flexible coronal model. For coronal modeling, {\sc pfsspy} \citep{Stansby2020} python package has been used in this work. The emphasis has been to achieve satisfactory results in modest computational time. This paper highlights the technical specifications and implementation of model to forecast and assess the ambient solar wind plasma properties at L1. Additionally, the paper demonstrates the prospect to compliment the in-situ measurements of ADITYA-L1.
\par
Aditya-L1 is India's first dedicated solar mission to be placed at a halo orbit around the first Lagrangian point of the Sun-Earth system. Solar Wind Ion Spectrometer (SWIS) and SupraThermal \& Energetic Particle Spectrometer (STEPS) are the two subsystems of Aditya 
Solar wind Particle EXperiment (ASPEX) payload on-board Aditya-L1. Brief details regarding the Aditya-L1 mission \citep{Seetha2017} and the ASPEX payload \citep{Janardhan2017, Goyal2018} have been reported elsewhere. The novelty of ASPEX is multi-directional, high cadence and proton-alpha separated measurements. In this work, the potential solar wind plasma measurements by SWIS have been simulated based on the modeled outputs to understand the directional variation of solar wind proton fluxes during the passage of SIR or Corotating Interaction Region (CIR).
\par
The paper has been organised in the following manner. The methodological description of framework is discussed in Section \ref{Section 2}. The model capability along with the comparison of results with observations at L1 is presented in Section \ref{Section 3}. The assessment of observation of ASPEX using SWASTi-SW has been done in Section \ref{Section 4}. A discussion on results, limitations, challenges and forthcoming projects is contained in Section \ref{Section 5}.

\section{SWASTi-SW} \label{Section 2}

    \begin{figure*}
       \centering
       \includegraphics[width = 0.9\textwidth]{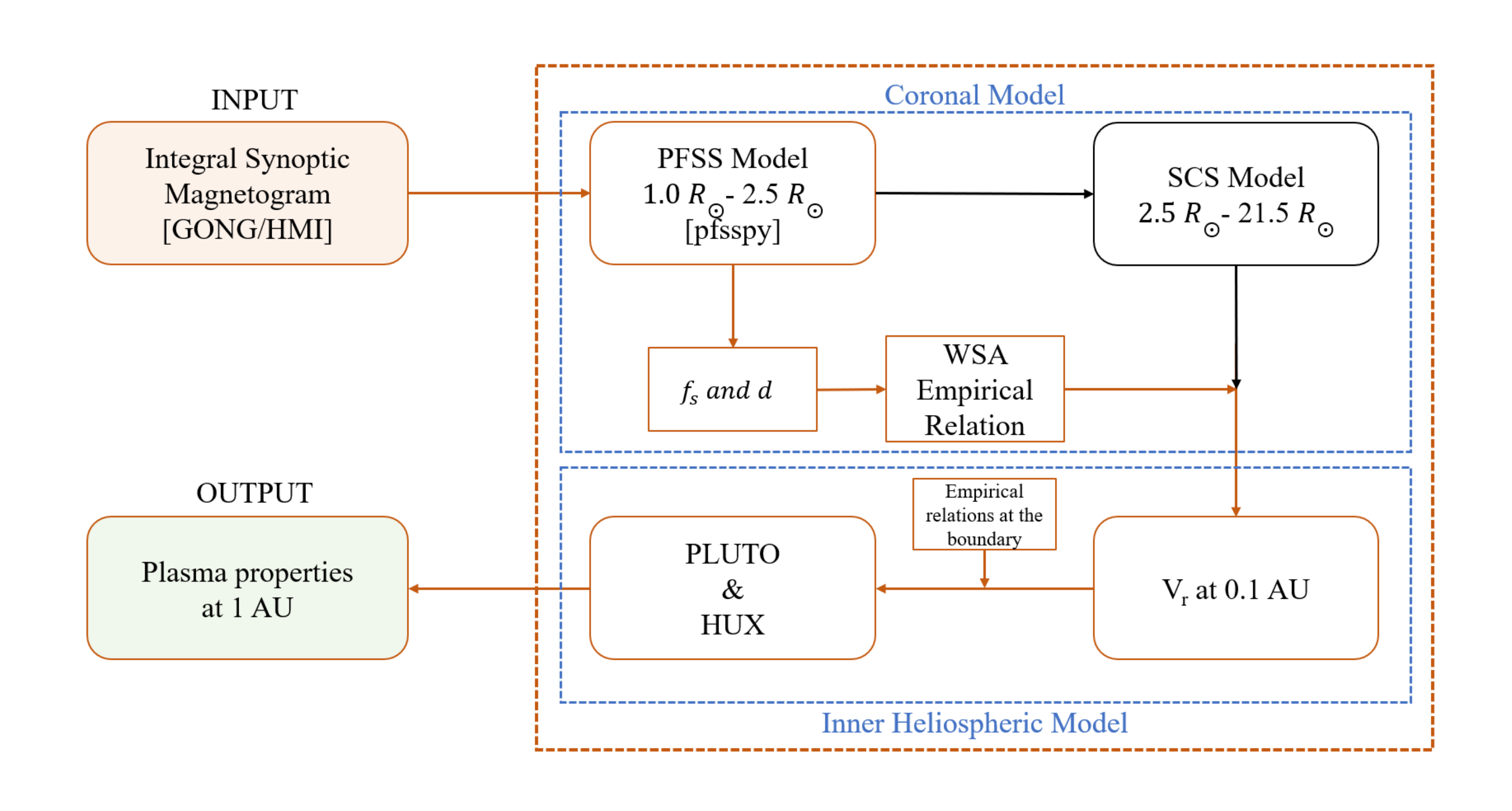}
       \caption{Process flow diagram of the proposed solar wind model showing the range of numerical models involved in the subdomains.}
       \label{fig:flowchart}
    \end{figure*}
 
   This numerical framework for forecasting and assessing the ambient solar wind is based on a well-established scheme which uses semi-empirical coronal model and physics-based inner heliospheric model. Figure \ref{fig:flowchart} shows the processes involved in SWASTi-SW, from photospheric magnetogram input to computing plasma properties in the inner-heliosphere. The spatial range of coronal domain goes from 1.0 $R_\odot$ to 21.5 $R_\odot$ (0.1 AU) and that of inner heliosphere from 0.1 AU to 2.1 AU. The mentioned scheme is now commonly used for the simulation of Sun-Earth connection, for example in ENLIL, SUSANOO, EUHFORIA, etc. Though a similar scheme is followed by these existing models, they differ in defining crucial parameters in both the subdomains. The details of sub domains of SWASTi-SW have been described in the following sub sections.

   \subsection{Sub-model for Corona}
   The primary aim of coronal domain is to provide inner boundary condition for the inner-heliospheric model, thereby the radial distance of this boundary ($R_{in}$), from the centre of the Sun, decides the range of coronal model. $R_{in}$ should essentially lie in the region where solar wind plasma becomes supersonic as well as super-Alfvenic. \citet{Goelzer2014} showed that this distance is correlated with the sunspot number and varies from 15 $R_\odot$ at solar minima to 30 $R_\odot$ at solar maxima. As in this work we have mainly focused around the solar minima regime, taking this distance to be 21.5 $R_\odot$ (i.e. 0.1 AU) is a physically suitable estimate. Therefore, the coronal domain's radial coverage extends to 0.1 AU, while its latitudinal and longitudinal coverage range from -90\textdegree$\,$ to 90\textdegree$\,$ and from 0\textdegree$\,$ to 360\textdegree$\,$, respectively, in heliographic Carrington frame.
   \par
   SWASTi-SW uses synoptic magnetogram as input and a modular approach for coupling Potential Field Source Surface (PFSS) \citep[]{Altschuler1969} and Schatten Current Sheet (SCS) \citep{Schatten1971} codes. This modular method facilitates an option of using PFSS alone or a coupled PFSS+SCS. Both versions rely on empirical relation of WSA model to calculate the solar wind speed profile at $R_{in}$.
   
   \subsubsection{Input Magnetogram}
   The only observational input in our model is full-disk photospheric magnetogram. Therefore it becomes important to carefully choose the suitable type of input magnetogram. In this work, we have used integral Carrington rotation (CR) synoptic maps provided by NSO-GONG (filename prefix: \textit{mrmqs}) and SDO-HMI (JSOC series: \textit{hmi.synoptic\_mr\_polfil\_720s}).
   The integral synoptic maps are calibrated by merging standard line-of-sight (LOS) maps and remapping into appropriate longitude in the Carrington frame. For each Carrington longitude of synoptic map, standard magnetograms near the central meridian, of that longitude, is averaged using a weighting factor (for example, GONG uses cosine$^{4}$(longitude)), for more details see \citet{Hill2018} and \citet{Scherrer2012}. The advantage of using integral CR synoptic maps is that each point along the X-axis (longitude) represents the location of Earth during that CR period, thereby providing the required input for studying the ambient solar wind at 1 AU in lesser computational time.
   \par
   Both GONG and HMI magnetograms provide magnetic fields at solar surface over linearly spaced grid points in longitude (X-axis) and equally spaced in sine(latitude) grid points in latitude (Y-axis). The used GONG magnetogram has resolution of $360\times180$ points whereas that of HMI is $720\times360$ points in phi-theta plane.

   \subsubsection{Potential field source surface model}
   In the lower solar corona, we have used PFSS model to solve for the global magnetic fields. It is based on a simple approach which exercises the uniqueness theorem of Laplace equation by assuming the electric current to be negligible. This approximation is reasonable in lower corona where plasma is force-free \citep{Gary2001} and most of the trans-equatorial fields lines are current-free \citep{Tadesse2014}, specially in quiet and weak active regions. Though more realistic, but complex, models exist see \citep[and references therein]{Wiegelmann2017}, observational tests \citep{Schrijver2003, Liu2008} and comparative studies e.g., \citep{Riley2006} depict that PFSS model is adequate for examining large-scale solar and heliospheric fields.
   \par
   PFSS solves Laplace equation from the solar surface (boundary condition provided by input magnetogram) to source surface (from where field is prescribed  to be radial). Traditionally, a spherical harmonic expansion approach is implemented (e.g., \citet{Altschuler1977, Hakamada1995, Nikolic2019}, WSA-ENLIL, EUHFORIA). However, this technique gives rise to ring-like patterns and is also sensitive to the choice of number of spherical harmonics \citep{Toth2011, Asvestari2019}. An iterative finite difference scheme has also been applied to solve PFSS \citep{VanDerHolst2010, Caplan2021} and it shows no signature of ringing effect and can be favoured over harmonic approach, specially near strong magnetic field regions. In this work, we have used {\sc pfsspy} which is based on the method of \citet{Ballegooijen2000}. {\sc pfsspy} code is a finite difference solver and hence allows the model to escape the ringing effect disadvantage occurring in spherical harmonic approach.
   \par
   {\sc pfsspy} is solved on a rectilinear grid which is equally spaced in ln(r), $\cos\theta$ and $\phi$ in spherical coordinates (r, $\theta$, $\phi$). We have used grid resolution of $100\times181\times361$ to solve for fieldlines from 1 $R_\odot$ to source surface radius ($R_{ss}$), which is 2.5 $R_\odot$ in our case. The magnetic fieldlines is traced in two sets. First, from inner boundary (1 $R_\odot$) to outer boundary (2.5 $R_\odot$) at the mentioned resolution, which gives coronal hole perimeter as both open and closed fieldlines are traced. Second, from source to solar surface at higher resolution of $100\times181\times N$ where $N$ is the number of hours in CR. In the later set, there are only open fieldlines originating from inside the coronal hole, whose perimeter is traced by the first set. This two set tracing approach provides greater resolution of fieldlines at the source surface.
               
    \begin{figure*}
       \centering
       \includegraphics[width=\textwidth]{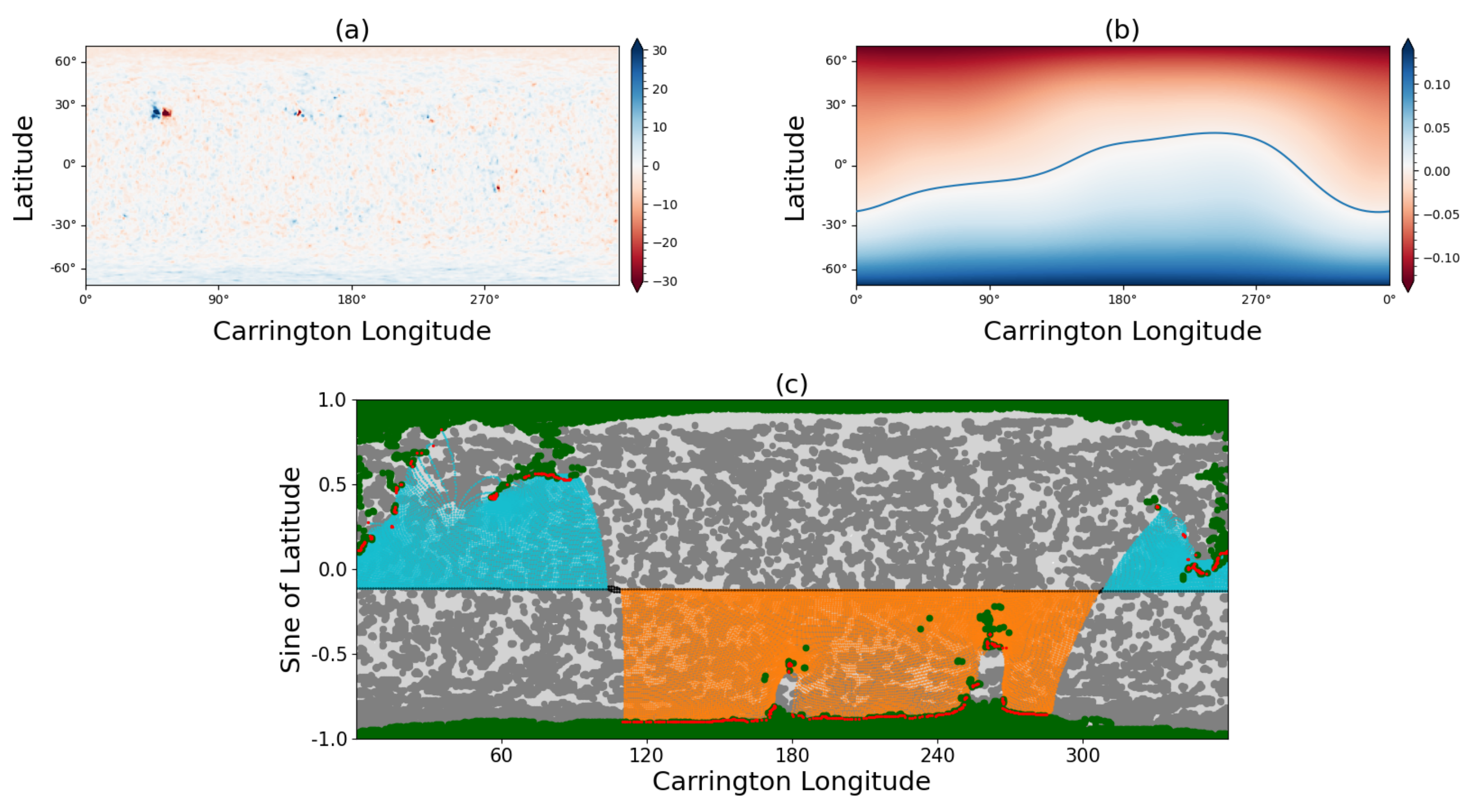}
       \caption{PFSS model results for CR2081. Here, (a) is computed radial magnetic field (Gauss) at 1 $R_\odot$, (b) is radial magnetic field (Gauss) output at 2.5 $R_\odot$ and, (c) is tracing of fieldlines from solar to source surface that will reach sub-Earth points. Here, cyan-orange and light-dark grey shows the open and closed fieldlines of opposite polarity. The green coloured area is coronal hole and red dots are the footpoint of open fieldlines. The horizontal black line is the location of Earth in heliographic coordinates for CR2081.}
       \label{fig:pfss}
    \end{figure*}
   
   Figure \ref{fig:pfss} shows the results of PFSS model, projected in $\theta-\phi$ plane. Subplot (a) shows the computed input radial magnetic field from the GONG magnetogram and (b) shows the radial magnetic field at the source surface ($R_{ss}$). In subplot (b), the blue line in the middle represents the magnetic polarity inversion line which effectively shapes the current sheet in heliosphere. Subplot (c) depicts the two set tracing approach of magnetic fieldlines. The tracing from 1 $R_\odot$ to $R_{ss}$ gives the region of closed fieldlines (light and dark grey) and the coronal hole boundary (the green area) at solar surface. Further, the tracing of open fieldlines (cyan and orange) from $R_{ss}$ to $R_\odot$ provides their foot points (red dots) inside the coronal hole area. The black line in the middle is the location of Earth for the CR2081. Hence, the figure illustrates the origin points of the fieldlines that will reach the Earth.
   
    \begin{figure*}[t]
       \centering
       \includegraphics[width=\textwidth]{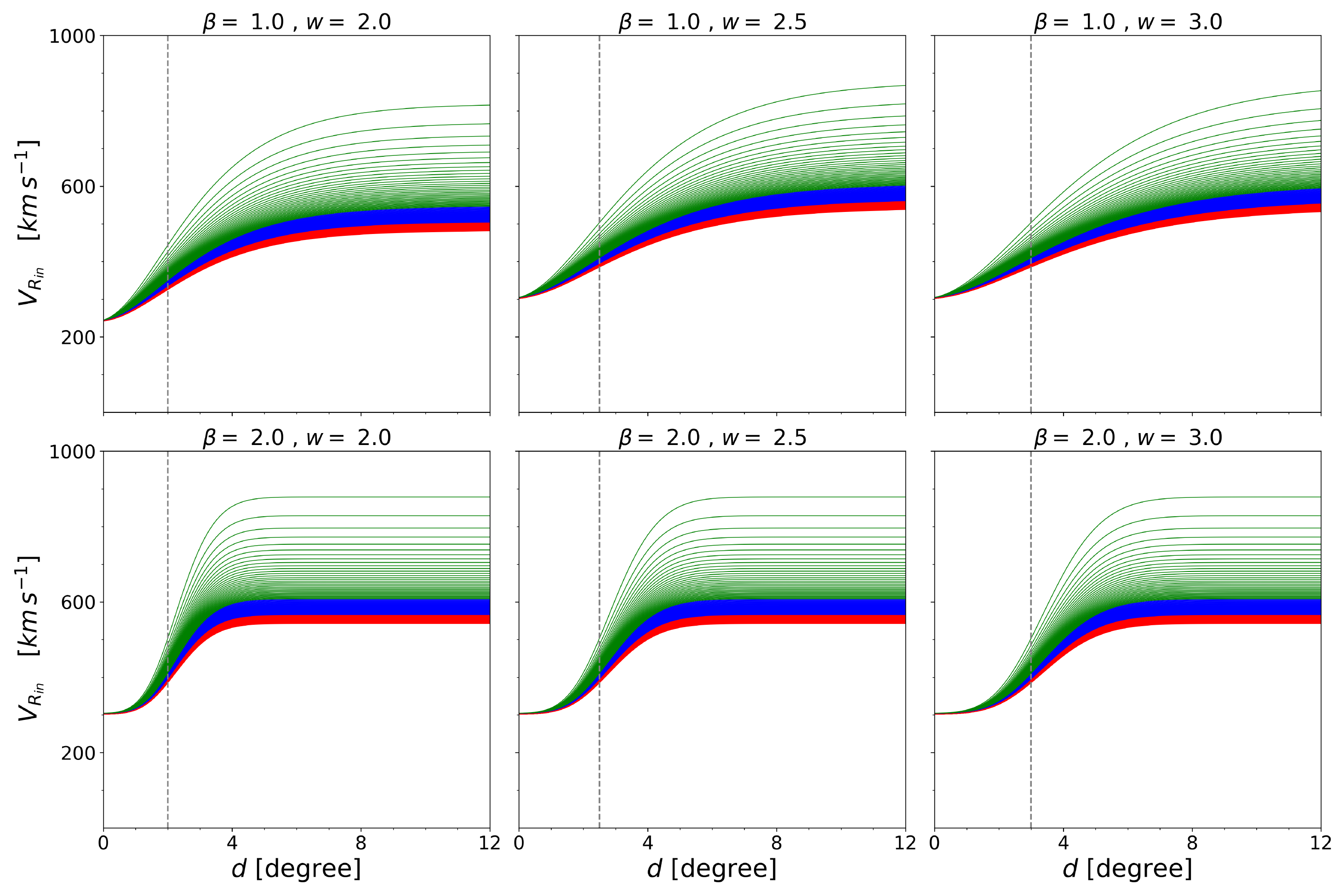}
       \caption{Variation of solar wind speed for the range of values of $d$ and $f_s$ based on different set of values of $\beta$ and $w$ while keeping the $V_{min}$ and $V_{max}$ constant (240 km/s and 725 km/s). $d$ varies from 0\textdegree $\,$ to 12\textdegree $\,$ and value of $f_s$ goes from 1 (green at top) to 99 (red at bottom) with increment of 1. The dotted vertical lines are placed corresponding to values of $w$. For $\beta=2$, $V_{R_{in}}$ becomes independent of $d$ for much smaller as compared to $\beta=1$. }
       \label{fig:d_vs_fs}
    \end{figure*}
   
   \subsubsection{Schatten current sheet model}
   In the upper corona (i.e. beyond $R_{ss}$), plasma beta becomes greater than unity \citep{Gary2001} and most of the fieldlines become almost radial, as plasma pressure starts dominating the dynamics. To incorporate the non-zero current region near the polarity inversion territory, \citet{Schatten1971} proposed SCS model. It requires solving another Laplace equation with inner boundary conditions given by PFSS and outer boundary extending to infinity, leading the fields to vanish. The inner boundary distance of SCS ($R_{scs}$) is usually taken less than $R_{ss}$ to avoid the kink formation at the interface. This technique results in improved solar wind structures at 1 AU sometimes, but most of the time it doesn't affect solar wind predictions \citep{McGregor2008}. In SWASTi-SW, the default setup is $R_{scs} = R_{ss}$, but it can be changed by the user at the run-time.
   \par
   There are two main advantages in using coupled PFSS+SCS over PFSS alone. It provides more realistic magnetic field values in the upper corona, and it facilitates more accurate fieldline tracing (by considering non-zero current region) in slightly more computational time.
   
   \subsubsection{Adapted WSA model}
   To provide the inner boundary conditions to MHD-based inner-heliospheric domain, we have used WSA model \citep{Arge2003}, which provides velocity profile at 0.1 AU for a given magnetic flux tube. Though there are many different forms of WSA solar wind relations, in SWASTi-SW we have used the following: 
   \begin{equation}\label{eq1}
       V_{R_{in}} = V_{min} + \frac{V_{max}}{(1+f_s)^{\frac{2}{9}}} \times \Bigg[\,\Bigg(1.0 - 0.8\,exp\, \Bigg(-\bigg(\,\frac{d}{w}\,\bigg)^{\,\beta}\,\Bigg)\,\Bigg)^{3}\,\Bigg] \quad km\,s^{-1}
   \end{equation}
    where, 
   \begin{equation}\label{eq2}
        f_s = \frac{R_\odot^2 \times B_r(R_\odot, \theta, \phi)}{R_{ss}^2 \times B_r(R_{ss}, \theta, \phi)}
   \end{equation}
   
   which is similar to equation (2) of \citet{McGregor2011}. In equation (\ref{eq1}), $V_{min}, V_{max},\, \beta$ and $w$ are independent parameters whereas, $f_s$ and d are areal expansion factor of flux tube and minimum angular separation of the foot-point from coronal hole boundary, respectively. A similar form of WSA velocity relation is also being used by other models (e.g., WSA-ENLIL, EUHFORIA, \citet{Narechania2021}) but each uses different set of values of independent parameters.
   \par
   The role of $w$ is to normalize the minimum angular distance of the flux tube foot-point from the open flux boundary ($d$) and $\beta$ controls the affect of this distance on solar wind relation ($V_{R_{in}}$). Whereas, $V_{min}$ and $V_{max}$ regulate the minimum and maximum value of $V_{R_{in}}$. Figure \ref{fig:d_vs_fs} shows the graphical representation of the functional form of $V_{R_{in}}$. Two main features are to be observed here. As the value of $d$ increases from 0\textdegree$\,$(i.e., fieldlines originating from close to the edge of coronal hole) the value of $V_{R_{in}}$ doesn't change much, regardless the value of $f_s$ and after a threshold value of $d$, $V_{R_{in}}$ depends only on $f_s$. Additionally, in between these two values, both $d$ and $f_s$ increases monotonically and contribute to $V_{R_{in}}$. And to properly use the capability of WSA model, speed empirical relation has to be dependent on both $d$ and $f_s$. Parameter $w$ regulates the value of $d$ from which second feature starts and $\beta$ determines the threshold value of $d$. The increase (decrease) in $V_{min}$ shifts the graphs upward (downward) and the variation in $V_{max}$ shifts the peak. Therefore, these four independent parameters in the empirical relation are critical in getting an accurate solar wind speed estimation at 0.1 AU.

    \begin{figure*}
       \centering
       \includegraphics[width = \textwidth]{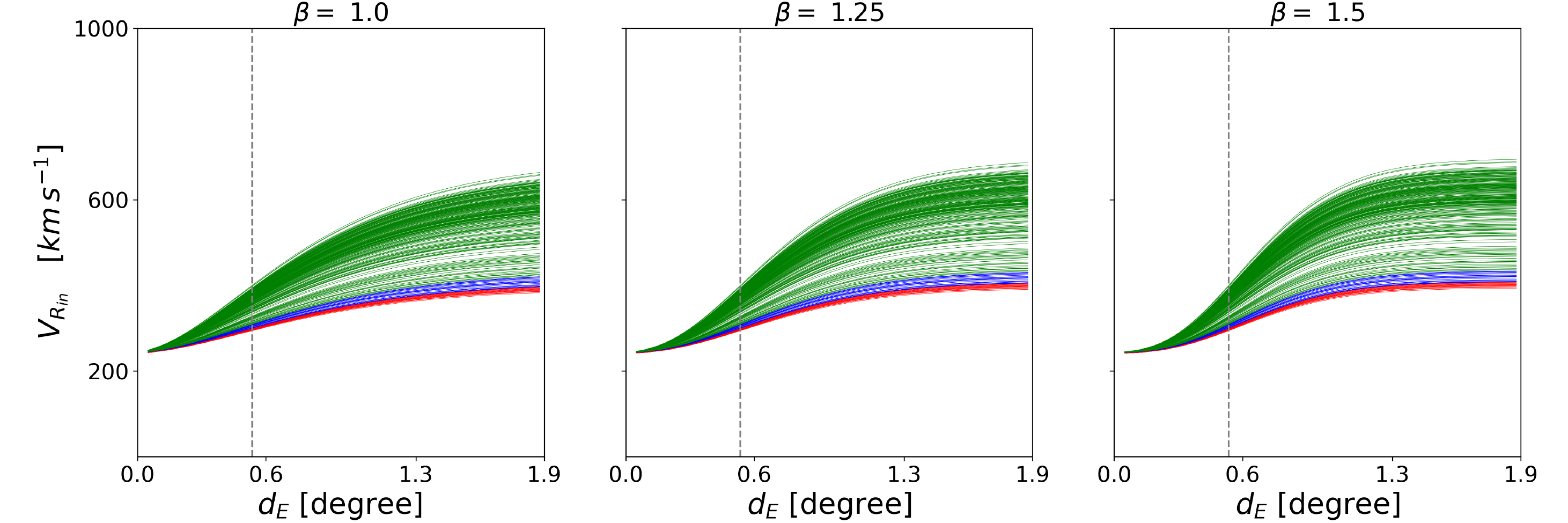}
       \caption{Graphical depiction of solar wind speed variation with different values of $\beta$ at 0.1\,AU for CR2081. Speed profiles are of those fieldlines that reaches sub-Earth points (location of Earth in Carrington heliographic coordinates). Here, $V_{min}=240\,km/s$, $V_{max}=725\,km/s$ and $w=0.54$ (median of $d_E$). For each fieldline, value of $d_E$ and $f_{s_E}$ are calculated from coronal model and the vertical dotted line represents the value of $w$. $f_{s_E}$ has been distributed in three equal bins (green, blue and red), with green showing the lowest (from top) and red showing the largest values (at bottom).}
       \label{fig:dmedian_vs_fs}
    \end{figure*}
   
   Out of four free parameters, the optimal value of $w$ is the most volatile, as for different grid resolutions the value of $d$ changes and so will $w$, to effectively normalize it. With an attempt to take a more generalized approach, we replaced the value of $w$ with the median of $d$. Keeping the focus on the fieldlines that reaches the location of Earth, we calculated $d$ and $f_s$ (now $d_E$ and $f_{s_E}$ for this case) values for only those flux tubes and checked the variability features in solar wind. As shown in Figure \ref{fig:dmedian_vs_fs}, this adapted method displays same kind of features as in Figure \ref{fig:d_vs_fs}. Moreover, by fixing the values of $V_{max}$ and $V_{min}$ (by default 725 km/s and 240 km/s for HUX run) we can presume, from Figure \ref{fig:dmedian_vs_fs}, that the optimal value of $\beta$ should lie near 1.0. But to find the optimal value of $\beta$ precisely, comparison of speed profile at first Lagrangian point of Sun-Earth system (L1) with observational data is required.
   \par
   For comparison with OMNIWeb data, we used Heliospheric Upwind eXtrapolation (HUX) model \citep{Riley2011}, for a range of values of $\beta$. HUX is a simple one dimensional upwind extrapolation technique which neglects the effects of magnetic field, gravity and pressure gradient. The HUX model gives very good match for speed results at 1 AU \citep{Riley2021} and that too in very less computational time. To find the optimal value of $\beta$ for a given CR, we calculated the solar wind profile at 1 AU ($V_{HUX}$) by varying $\beta$ from 0.75 to 1.75 in 20 equal steps and compared it with observed data ($V_{OBS}$). For this initial study we restricted the range to $\pm{0.5}$ around the most used value of $\beta$ i.e., 1.25 e.g., \citep{VanDerHolst2010, Riley2015, Pomoell2018, Narechania2021}. To statistically evaluate the best fit, we used Pearson correlation coefficient ($cc$), root mean square error ($rmse$) and normalized difference of standard deviation of $V_{HUX}$ and $V_{OBS}$ ($nsd$). The best match was decided on the basis of a score ($\Sigma$ in equation \ref{eq3}) giving equal weightage to all three, lower $\Sigma$ value signifies better match.
   
   \begin{equation}\label{eq3}
         \Sigma = (1-cc)^{\,2} + \bigg(\frac{rmse}{100}\bigg)^{\,2} + nsd^{\,2}
   \end{equation}
   
\begin{figure*}[t]
       \centering
       \includegraphics[width = \textwidth]{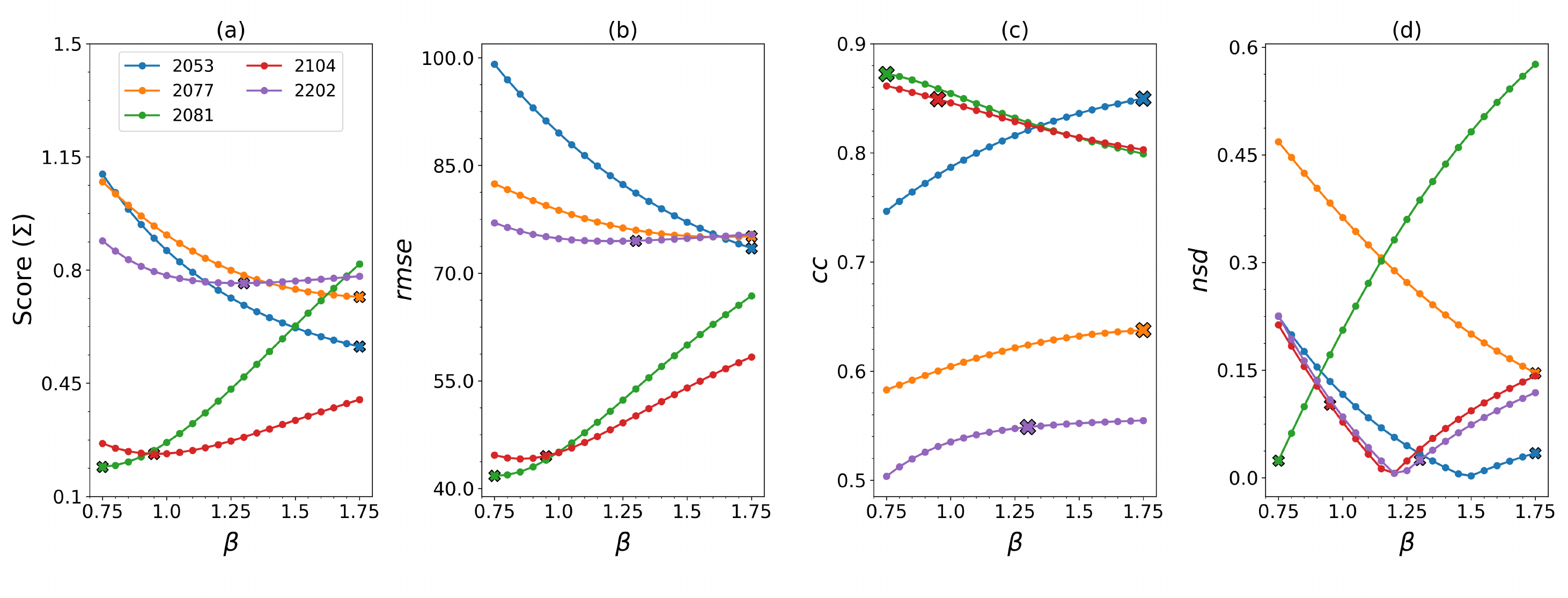}
       \caption{Plots showing the statistical results for variation of $\beta$ from 0.75 to 1.75 in 20 equal steps, for five selected CRs. On the basis of the minimum score value, optimum $\beta$ value has been deduced. The chosen optimum for each CR, has been marked with 'x', whose values have been mentioned in Table \ref{tab:optimal_beta}. All subplots have common plot legend, shown in subplot (a).}
      \label{fig:optimum_beta}
    \end{figure*}

   We selected five CRs for comparison, near the solar minima region and in the absence of halo CMEs to properly capture the features of ambient solar wind. Considering that the accuracy of GONG magnetogram has degraded since 2013, specially in polar regions \citep{Nikolic2019}, we focused more on increasing phase of Solar Cycle 24. The optimal value of $\beta$ for selected CRs, based on HUX, have been listed in Table \ref{tab:optimal_beta} and variations are shown in Figure \ref{fig:optimum_beta}. The subplot \ref{fig:optimum_beta}(a) shows the variability of $\Sigma$ value for all CRs. As $\beta$ increases from 0.75 to 1.75, the change in $\Sigma$ value is lesser for CR2077, CR2104 and CR2202 (decline of $<$0.3) whereas, CR2053 and CR2053 show greater deviation (incline of $>$0.4). The optimal $\beta$ value lies in the middle of the chosen range for CR2104 and CR2202, on the contrary, it lies at the boundary for CR2053, CR2077 and CR2081. That means, for the later three CRs the real optimum value can be located outside the chosen range of $\beta$. But at their optimum value, the slope of their plots have become almost zero (fig. \ref{fig:optimum_beta}(d)). This indicates that no significant reduction in $\Sigma$ value will occur with further change of $\beta$ and therefore the chosen range is adequate.
   
    \begin{table}
    \caption{Statistical results of selected CRs and their optical value of $\beta$.}
    \centering
    \begin{tabular}{l c c c c c}
    \hline
    {CR} & {Optimal value of $\beta$} & {$cc$} & {$rmse$} & {$nsd$} & {$\Sigma$}           \\
    \hline
    2053        & 1.75          & 0.85        & 74.01        & 0.04         & 0.57    \\
    2077        & 1.75          & 0.64        & 75.20        & 0.14         & 0.72    \\
    2081        & 0.75          & 0.87        & 41.39        & 0.03         & 0.19    \\
    2104        & 1.0           & 0.85        & 44.51        & 0.09         & 0.23    \\
    2202        & 1.3           & 0.54        & 74.35        & 0.02         & 0.77    \\
    \hline
    \end{tabular}
    \label{tab:optimal_beta}
    \end{table}
   
   In Figure \ref{fig:optimum_beta}, root mean square error (rmse) has monotonically increased with $\beta$ for CR2104 and CR2081, decreased for CR2053 and remained almost constant for CR2077 and CR2202. The Pearson correlation coefficient (cc) became better with increased value of $\beta$ for CR2053, CR2077 and CR2202, whereas $cc$ value decreased for CR2081 and CR2104. The normalized standard deviation (nsd) first increased and then decreased for CR2053, CR2077 and CR2202. And in case of CR2081 (CR2104) the nsd value strictly increased (decreased) with $\beta$. 
   \par
   There is no noticeable pattern among the CRs in plots of Figure \ref{fig:optimum_beta}, but different optimum $\beta$ values for different CRs emphasises the influence of $d$ parameter. Higher $\beta$ value implies greater influence of $d$ on speed profile, as compared to its lower value. Therefore it can be inferred that in the WSA relation (equation \ref{eq1}), dominance of $d$ is greater in CR2053 ($\beta$=1.75) as compared to CR2081 ($\beta$=0.75).
   \par
\begin{figure*}
       \centering
       \includegraphics[width = \textwidth]{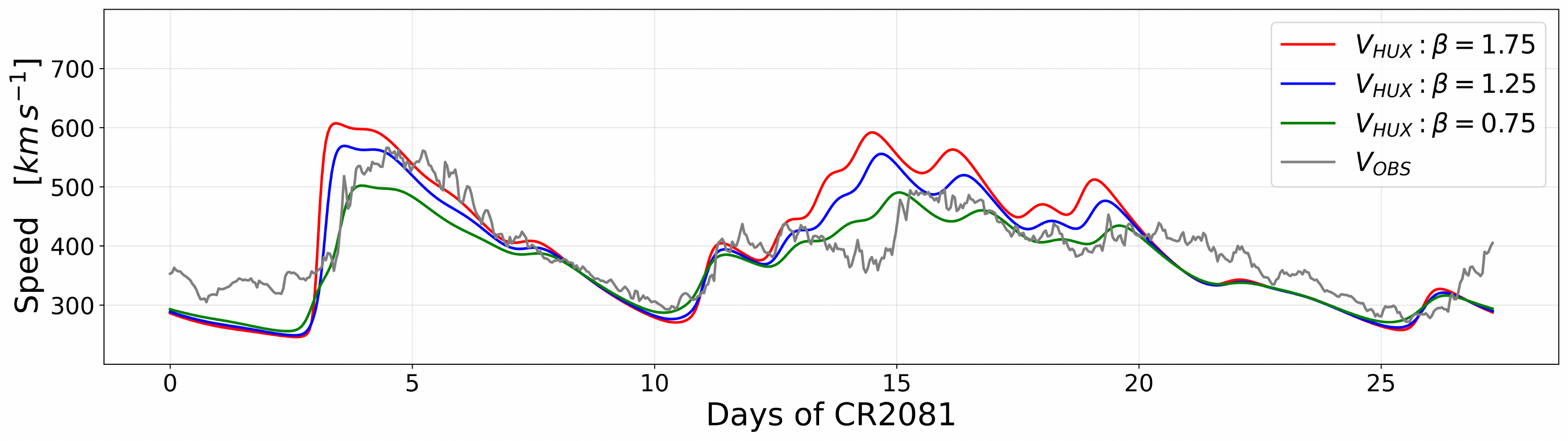}
       \caption{Effect of value of $\beta$ on speed profile at 1 AU. The difference in speed is greater at peaks and lesser at base. The speed profile have smoother peaks for low value of $\beta$. }
       \label{fig:beta_2081}
    \end{figure*}

   As expected, Figure \ref{fig:beta_2081}  demonstrates that even small variations in $\beta$ have significant impacts on the speed profile, particularly at peaks, where the value of $d$ is high. As the $\beta$ value is increased from 0.75 to 1.25 and 1.75, the peak shifts upward and it's fluctuations in that region also increases. The same trend will continue till the threshold value is reached and then no further change will happen, as speed profile becomes independent of $d$ (see fig. \ref{fig:dmedian_vs_fs}).
  
    \subsection{Sub-model for Inner Heliosphere}
    The fundamental purpose of MHD-based inner heliospheric model is to determine plasma properties in inner heliosphere by taking the input from coronal model. The veracity of this domain depends on accurate initial boundary conditions which vastly depends on the speed profile derived from WSA relation. More details has been prescribed in the following subsections.
            
    \subsubsection{MHD Setup}
    The inner heliospheric model is based on {\sc pluto} code \citep{Mignone2007}, which is built on Godunov-type schemes to integrate a set of conservation laws using finite volume or finite difference approach. In the current version of SWASTi-SW, ideal MHD module of {\sc pluto} is used on a uniform static grid in spherical coordinates. The following set of conservative equations are solved using finite volume method:

    \begin{align}
        \pdv{\rho}{t} \quad + \quad \nabla \cdot (\rho \textbf{v})\quad = \quad0 \label{eq4}\\
        \pdv{\textbf{m}}{t}  \quad + \quad \nabla \cdot \bigg[\textbf{mv} - \textbf{BB} + \bigg( p + \frac{\textbf{B}^{2}}{2}\bigg)\bigg] \quad = \quad \rho\textbf{g} \label{eq5} \\
        \pdv{\textbf{B}}{t} \quad - \quad \nabla\,\times \,(\textbf{v}\times\textbf{B}) \quad = \quad 0 \label{eq6}\\
        \pdv{E_t}{t} \quad + \quad \nabla \cdot \bigg[\bigg(\frac{\rho \textbf{v}^{2}}{2} + \frac{\gamma p}{\gamma-1} \bigg)\,\textbf{v} + \textbf{B} \times (\textbf{v}\times\textbf{B})\bigg] \quad = \quad\textbf{m.g} \label{eq7}
    \end{align}
    
    where $\rho$ is mass density, \textbf{v} is velocity, \textbf{m} is momentum density ($\rho$\textbf{v}), \textbf{B} is magnetic field, $p$ is isotropic thermal pressure, \textbf{g} is gravitational acceleration (-$\frac{GM_{\odot}}{r^2}$), $E_t$ is total energy density and $\gamma$ (=5/3) is specific heat ratio of solar wind plasma. The above equations are solved in the Stonyhurst Heliographic frame. The coordinates of this frame can be converted to Heliocentric Earth Equatorial (HEEQ) frame by mere spherical to cartesian coordinate transformation \citep{Thompson2006}. To incorporate the solar rotation in this frame, inner radial boundary (the whole spherical slice at $R_{in}$) is rotated with constant angular speed with respect to the computational grid. The rotational time period (TP) remains constant for a specific CR and can have values from 27.21 to 27.34 days \citep{Thompson2006}, depending on the location of Earth in its orbit. For example, value of TP for CR2053 (starting from February, 2007) is 27.34 days whereas, value of TP for CR2081 (starting from March, 2009) is 27.30 days. And the corresponding centrifugal and Coriolis terms have been neglected due to their trifling share.
    \par
    The MHD domain range goes from 0.1 AU to 2.1 AU in radial, -60\textdegree $\,$ to 60\textdegree $\,$ in latitudinal and 0\textdegree $\,$ to 360\textdegree $\,$ in longitudinal direction, having $150\times120\times360$ grid resolution respectively. With the motivation to keep the computational time reasonable, we opted for simple numerical methods involved in a typical time step cycle in {\sc pluto}. For each step we chose RK2 time stepping algorithm, $2^{nd}$ order TVD linear reconstruction scheme and HLLC Riemann solver. And to ensure the divergence free condition, we selected Powell's 8 wave formulation \citep{Powell1994}.
    
    \subsubsection{Boundary Specification}
       
    \begin{figure*}
       \centering
       \includegraphics[width = \textwidth]{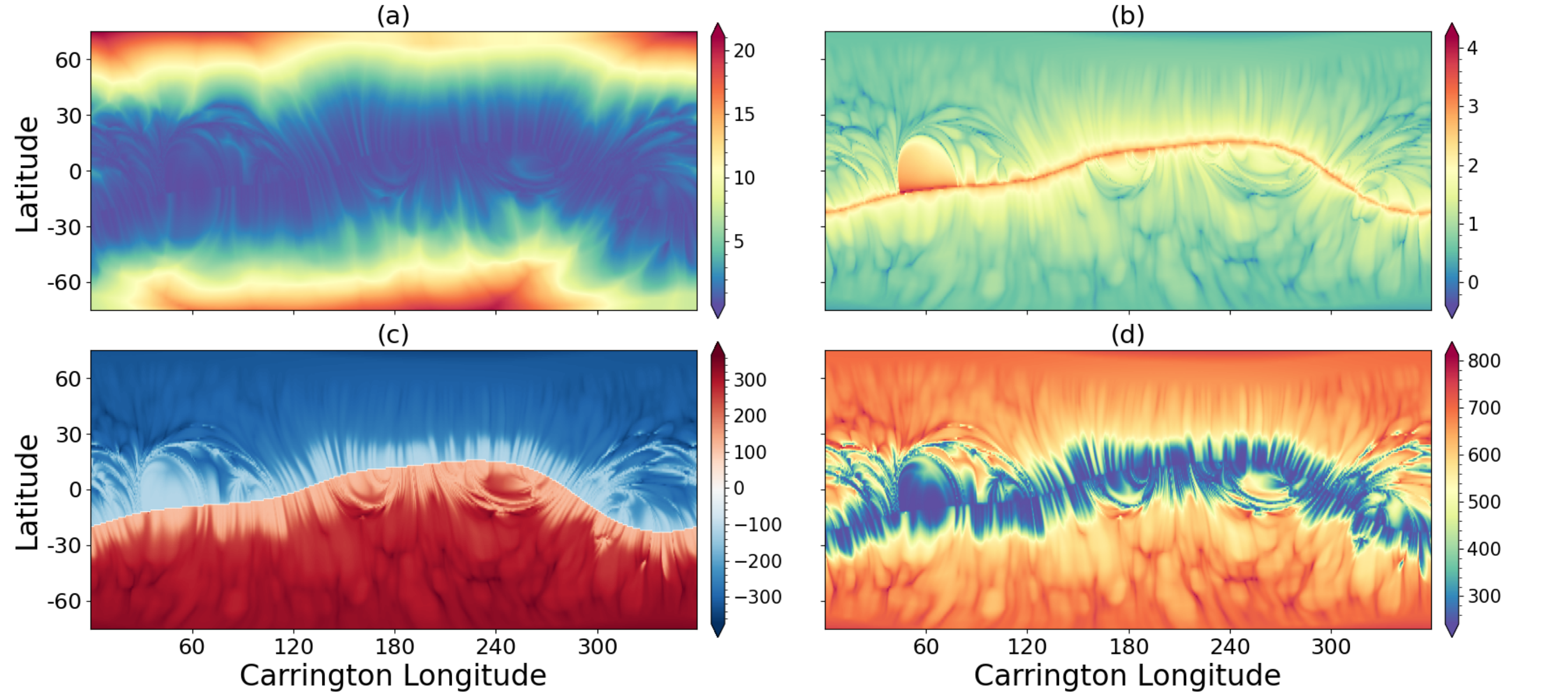}
       \caption{Plots for CR2081: (a) $d$ parameter in degree, (b) $\log_{10}(f_s)$, (c) radial magnetic field [nT] at $R_{in}$, derived from equation \ref{eq10}, and (d) solar wind speed [km/s] profile as input for MHD domain.}
       \label{fig:d_fs_Br_Vr}
   \end{figure*}
   
    The coronal domain provides the speed profile (radial component of \textbf{V}, $V_r$), for each flux tube,  at the inner boundary of MHD domain ($R_{in}$). For forecasting purpose, the default values of parameters in the adapted WSA relation, for MHD run, are: $V_{min}=250\,km\,s^{-1}$, $V_{max}=650\,km\,s^{-1}$, $w$ = median of $d_E$ and $\beta$ = 1.25, but in this paper $\beta$ values listed in table \ref{tab:optimal_beta} has been taken for assessment. Here, the value of $V_{max}$ is 75 $km\,s^{-1}$ less than the value that was used in HUX. A decreased value of $V_{R_{in}}$ has been applied to retaliate the affect of additional acceleration in MHD domain \citep{McGregor2011}. As the coronal domain doesn't include solar rotation, the speed profile is rotated in the longitudinal direction by angle $\alpha$:

    \begin{align}
        \alpha = 5\text{\textdegree} + \bigg(\frac{2\pi}{\text{TP}}\bigg)\,\bigg(\frac{20.5 R_\odot}{(V_{R_{in}})_{min}}\bigg) 
    \end{align}
    
    where $(V_{R_{in}})_{min}$ is minimum value of $V_{R_{in}}$.\\
    The values of other plasma properties at $R_{in}$ are derived from the following empirical relations:
    
    \begin{align}
        \label{eq9} n  &=  n_0 \, \bigg(\frac{V_{max}}{V_r}\bigg)^{\,2} \\
        \label{eq10} B_r  &=  \text{sgn} (B_{corona})\,B_0\,\bigg(\frac{V_r}{V_{max}}\bigg) \\
        \label{eq11} B_{\phi}  &=  -\,B_{r}\,\sin{\theta}\,\bigg(\frac{V_{rot}}{V_r}\bigg)
   \end{align}
    
   where $n$ is plasma number density, $n_0$ = 300 $cm^{-1}$, $B_r$ and $B_{\phi}$ are radial and azimuthal components of \textbf{B}, $B_0$ = 300 nT, $V_{rot}$ is rotating speed of inner boundary corresponding to TP. Here, $n_0$ and $B_0$ refer to number density and magnetic field values of fast solar wind, respectively. The thermal pressure, $p$ has been kept constant at $R_{in}$ at 6.6 nPa. The meridional and azimuthal components of velocity ($V_\theta$ and $V_\phi$) are assumed to zero at $R_{in}$. The empirical relations (equation \ref{eq9}, \ref{eq10} and \ref{eq11}) are similar to those used in \citet{Odstrcil2003} and \citet{Pomoell2018}. The coronal models usually underestimate the heliospheric magnetic flux, i.e., a significant part of magnetic flux goes undiscovered \citep{Linker2017}. Therefore, implementing an empirical relation, based on properties of fast wind and speed of that flux tube, sidesteps this problem.
   \par
   Figure \ref{fig:d_fs_Br_Vr} exhibits the input of MHD model for CR2081 at 0.1 AU. Plots (a) and (b) are of $d$ and $f_s$ (in logarithmic scale) which are used to compute the solar wind speed (plot (d)) using the adapted WSA relation and further, speed profile is used to evaluate radial magnetic field (plot (c)) using equation \ref{eq10}. Though the latitudinal range of MHD domain goes from -60\textdegree\ to +60\textdegree, the quantities at 0.1 AU are calculated for a wider range.
   \par
   At the outer boundary, the radial direction is set to outflow condition, i.e., zero gradient across the boundary whereas,  latitudinal and longitudinal direction are reflective and periodic, respectively on both sides of the computational domain. In reflective boundary condition, the variables are symmetrized across the boundary.
   \par
   It is worth mentioning that due to the usage of above empirical relations, the only difference left between coupled PFSS+SCS and PFSS alone is fieldline tracing technique. SCS provides more realistic tracing by incorporating the non-zero current in the current sheet region and as a result latitudinal value at $R_{in}$ differs from pure radial (zero current) extrapolation, which is used in case of PFSS alone approach. Usually, this latitudinal difference is less than 1\textdegree$\,$ for most of the fieldlines and its effect can truly be noticed only when model resolution is better than 1\textdegree. Therefore, for lower resolution setup, PFSS+SCS and PFSS alone won't show significant difference. In this work, we have used the PFSS alone procedure by bypassing the SCS model.

\section{Solar Wind Forecasting and Assessing Capabilities}\label{Section 3}

The motivation of this work is to develop a solar wind forecasting model which could run on a personal workstation in reasonable computational time. The model setup was formulated accordingly and ran on a workstation having 48 cores which computes the final result for mentioned resolution in approximately 6.5 to 9.5 hours. To validate the model output, we introduce the initial results of the current version of SWASTi-SW for selected CRs. The chosen CR number (starting month) are: CR2053 (Feb, 2007), CR2077 (Nov, 2008), CR2081 (March, 2009), CR2104 (Nov, 2010) and CR2202 (March, 2018). The sample CRs are taken in a way to cover regions around the minima of solar cycle. The first CR resides in the minima region of descending phase of solar cycle 23 and the later three CRs belong to the ascending phase of solar cycle 24, starting from its minima whereas, the fifth CR corresponds to minima of descending phase.

    \begin{figure*}[t]
    \begin{interactive}{animation}{Fig8_animation}
    \centering
    \includegraphics[width = 0.93\textwidth]{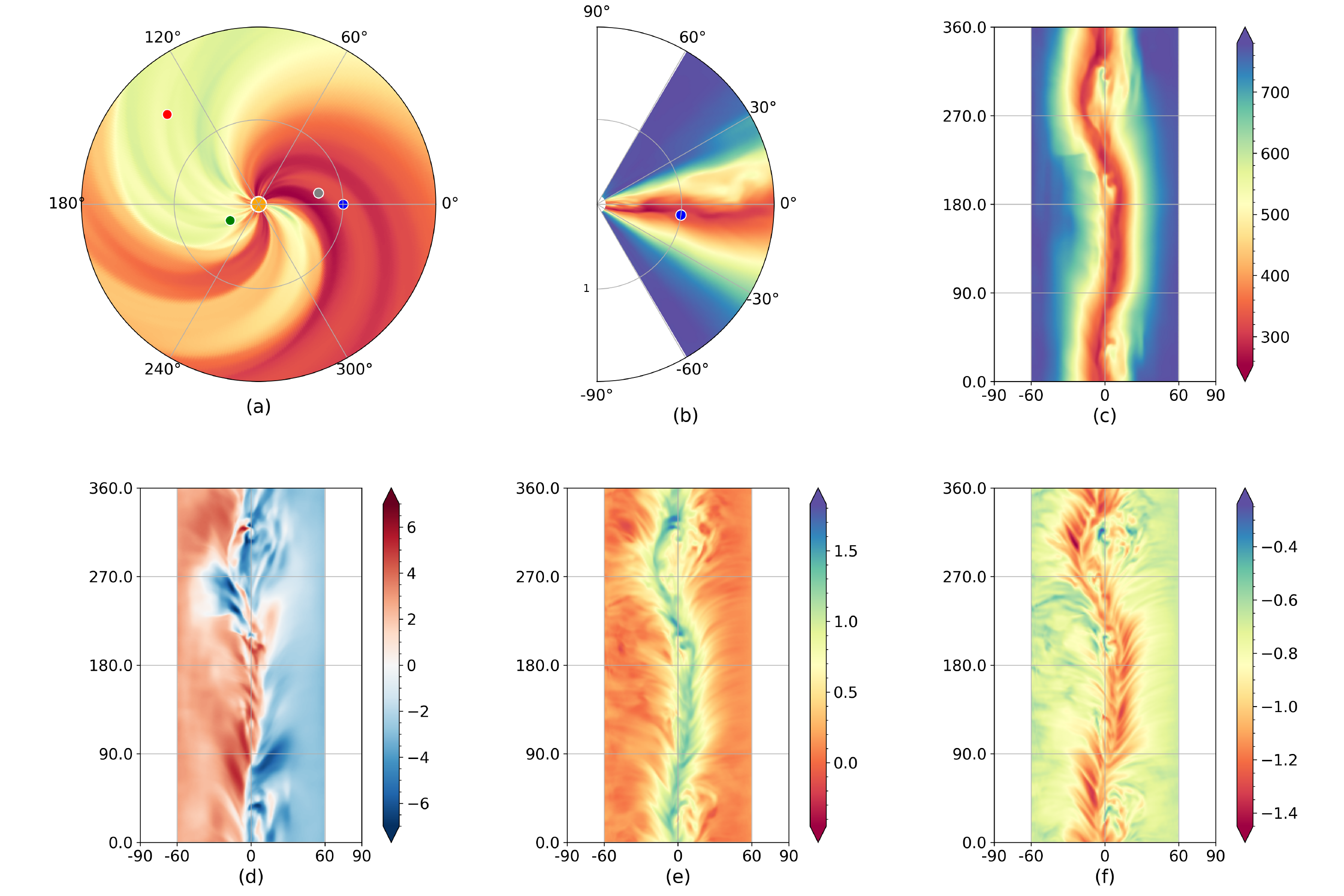}
    \end{interactive}
    \caption{Snapshots of output of the inner heliospheric model. The results are for CR2081. Here (a), (b) and (c) are the radial velocity plots whereas, (d) is plot of radial magnetic field (nT), (e) is of proton density (N$_p$cm$^{-3}$) in logarithmic scale and (f) is of proton temperature (MK) in logarithmic scale. (a) plot is in $r-\phi$ plane at earth's latitude location, (b) is in $r-\theta$ plane at 0\textdegree \ longitude, (c), (d), (e) and (f) are in $\theta-\phi$ plane at 1 AU. The blue dot at 1AU in (a) and (b) is the location of earth at the starting of CR2081. The green, grey and red dots in plot (a) denotes Mercury, Venus and Mars whereas, orange dot at the center highlights the Sun. This figure is available as an animation.}
    \label{fig:PLUTO_output}
    \end{figure*}
    
The spatial domain of MHD region goes up to 2.1 AU, covering the region of Mercury, Venus, Earth and Mars. A snapshot of the output is shown in Figure \ref{fig:PLUTO_output} where, plots (a), (b) and (c) are of radial velocity in different planes and (d), (e) and (f) are plots of radial magnetic field, proton density and proton temperature, respectively. Plots (c), (d), (e) and (f) are in $\theta-\phi$ plane at 1 AU and density and temperature have been shown in logarithmic scale to display the structure clearly. A heliospheric current sheet, where the polarity of magnetic field changes, can be observed in the middle of plot (d). The fieldlines near the current sheet region originates from the edge of coronal hole, therefore solar wind speed must be low in this region. And slow solar wind have higher density, which in turn leads to lower thermal temperature. As expected, plots (c), (e) and (f) also have current sheet structure with lower values of speed and temperature whereas, higher value of density near the heliospheric current sheet region.
\par
To compare our model results with observation, we used per hour averaged solar wind magnetic field and plasma data from OMNIWeb database. We interpolated the model output from 360 data points to number of hours in CR (say, N) for comparability. The model output and performance analysis are in the following subsections.
    
    \subsection{Plasma Properties at L1}
\begin{figure*}
       \centering
       \includegraphics[width = \textwidth]{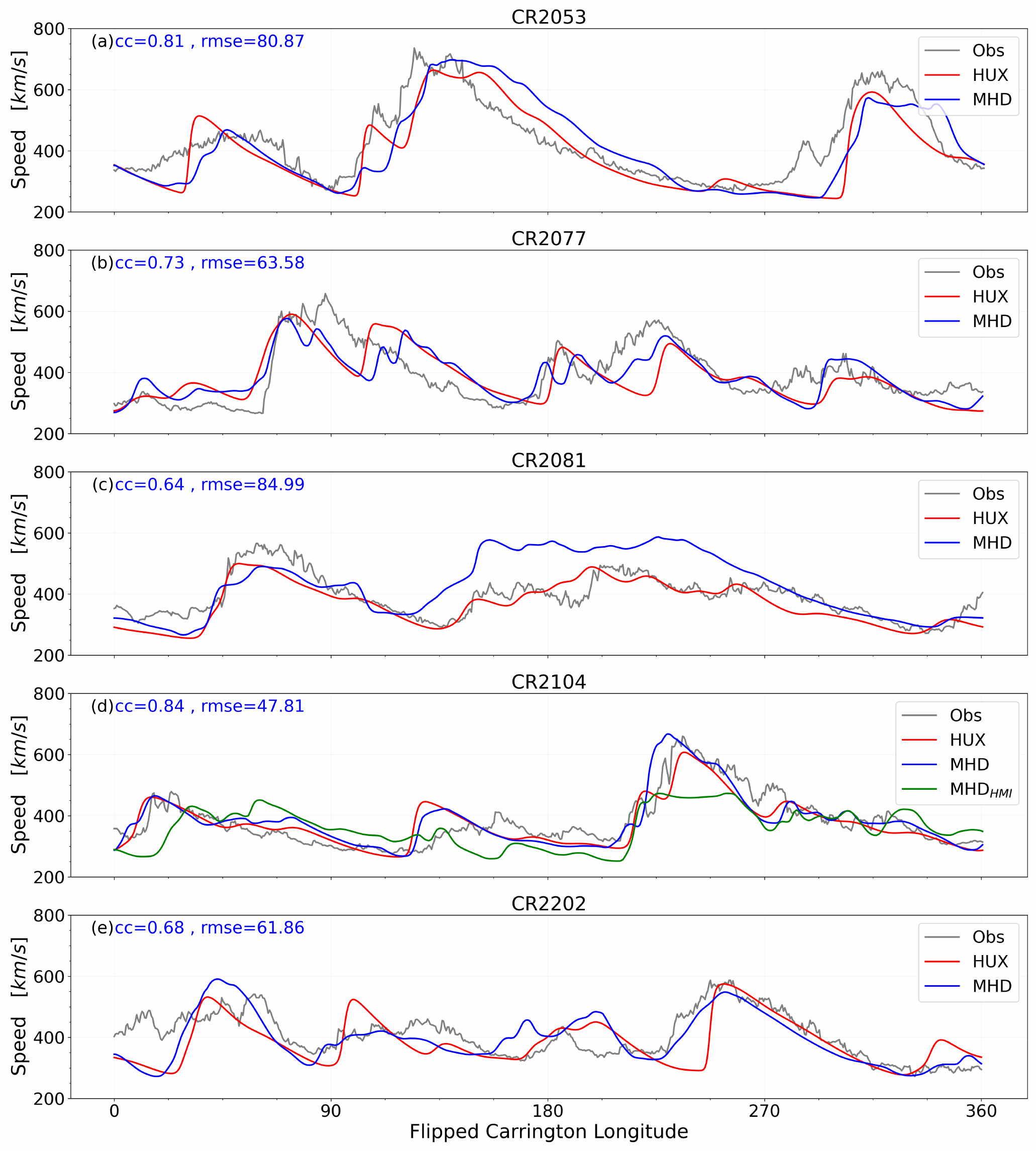}
       \caption{Plasma speed profile at L1 for mentioned CRs. For CR2053, CR2077, CR2081 and CR2022 MHD (using GONG magnetogram) and HUX results are plotted along with the per hour averaged OMNI data. For CR2104, additionally MHD$_{HMI}$ (using HMI magnetogram) have been plotted along with MHD (using GONG), HUX and OMNI data. The Pearson correlation coefficient (cc) and root mean square error (rmse) have been shown for MHD$_{GONG}$ results, which are in blue.}
       \label{fig:Vel_allCRs}
   \end{figure*}

\begin{figure*}[t]
       \centering
       \includegraphics[width = \textwidth]{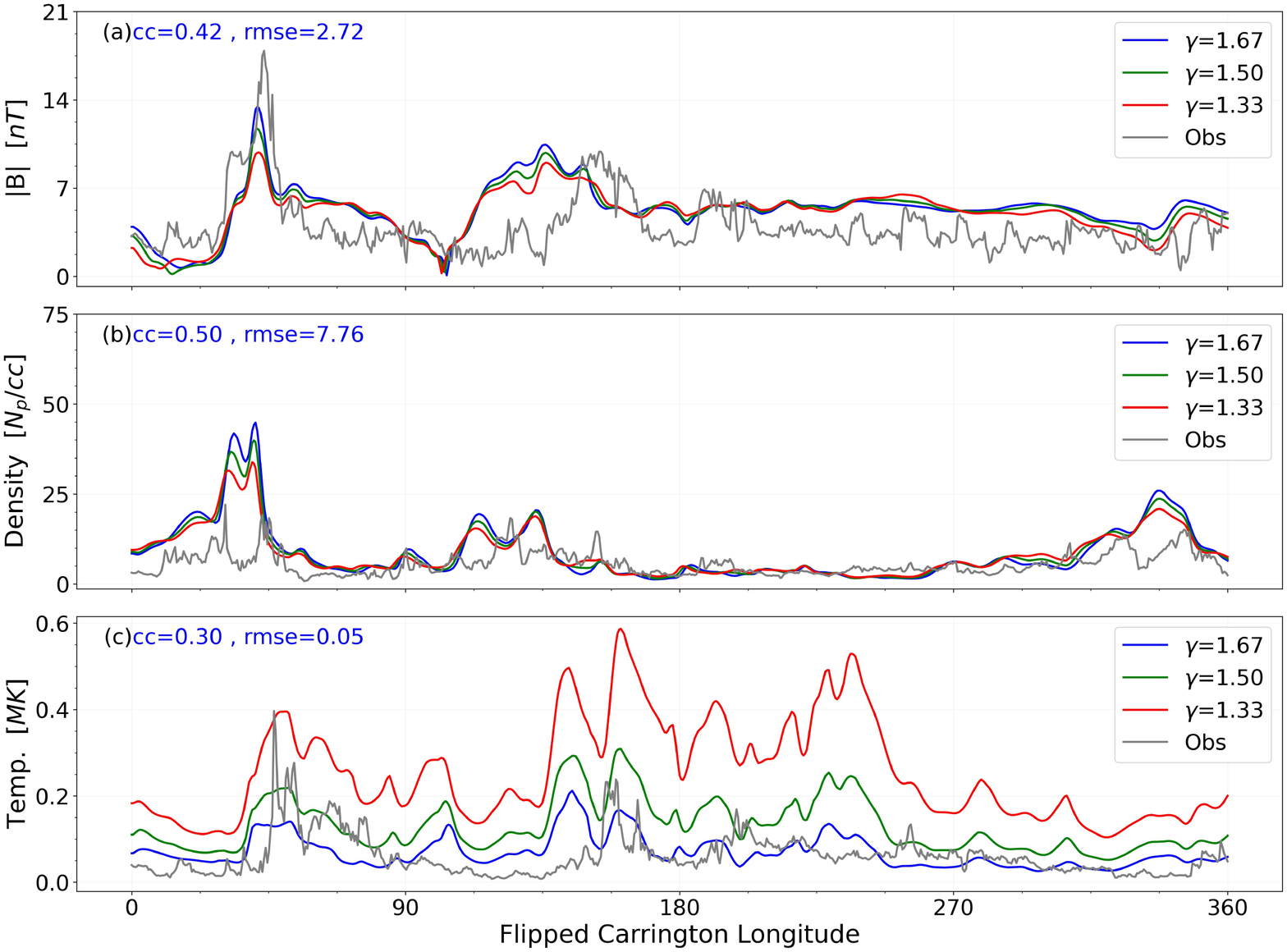}
       \caption{Plots showing the comparison of plasma properties with the OMNI data at L1 for different values of specific heat ratio, $\gamma$. The results shown here are for CR2081 and the value used in this work is shown in blue ($\gamma=5/3$). The Pearson correlation coefficient (cc) and root mean square error (rmse) are shown for MHD$_{GONG}$ with $\gamma$=5/3.}
       \label{fig:plasma_2081}
    \end{figure*}

    Figure \ref{fig:Vel_allCRs} shows plasma speed for all the selected CRs at L1, using GONG (all CRs) and HMI (one CR) magnetograms. In this Figure, the results of our MHD model and HUX technique has been compared with OMNI one hour averaged data. Since, the observed plasma speed is mainly radial, it has been compared with the radial velocity of the model. The X-axis is flipped Carrington longitude where 0\textdegree\ and 360\textdegree\ signify the start and end time of CR, respectively. Technically, the Carrington longitude starts from 360\textdegree\ and ends at 0\textdegree\ therefore, it has been flipped to keep the CR starting time and 0\textdegree\ as origin on the left side. The comparison with other plasma properties are in Figure \ref{fig:plasma_2081} where additionally, magnetic field magnitude, number density and proton temperature has been shown. 
    
    The speed profile output of SWASTi-SW for CR2053 has successfully captured the observed structure ($cc$ = 0.81 and $rmse$ = 80.87 km/s), which has two local sideward peaks and a global peak at the centre in Figure \ref{fig:Vel_allCRs}. Though the positions of sideward peaks have matched well, the global peak seems to be slightly shifted to the right by a few hours.
    \par
    In CR2077, the leftmost peak has bifurcated and become broader, but the positions of rest of the structure show a decent match. A considerable difference in the slope of HUX and MHD profiles is visible around 210\textdegree\ Carrington longitude, where physics-based result gives a better match with the observed value. As HUX technique neglects the effects of pressure gradient, the gradual increase in solar wind speed was not manifested which MHD approach successfully produced. Considering the fact that HUX extrapolates in only one direction, latitudinal flow of plasma could also be playing a role.
    \par
    The speed is higher in the middle region for CR2081, but the overall pattern is the same. The HUX result has a better match in this case. Whereas, in the case of CR2104, GONG magnetogram results are having a good match for both HUX ($cc$ = 0.86) and MHD ($cc$ = 0.84). They have accurately captured the whole structure except for a minor peak in the middle. But HMI result for CR2104, MHD ($cc$ = 0.57) is not matching that well, it has failed to reproduce the global maxima, near 240\textdegree, accurately.
    \par
    For CR2202, the same slope difference feature is visible around 240\textdegree\ as it was in CR2077 at 210\textdegree. Again the physics-based result is able to capture the observed form whereas, HUX is showing a very sharp peak at that position. An abrupt fall at the beginning of profile can be noticed, as compared to the observed one. A possible reason could be the cyclic nature of the model input, which enforces the same value at the end and beginning of the cycle.
    
    Figure \ref{fig:plasma_2081} shows the plots of other plasma properties for CR2081. The magnetic field output and observation data has the same order of magnitude as it varies under 20 nT. The overall pattern is also showing the same trend. But, the one hour averaged OMNI data has very rapid fluctuations which is missing in the interpolated model output. On the other hand, density plot is always higher than the observed one, indicating that the model is overestimating the plasma number density at L1. Whereas, the temperature plot's peaks are always lower, implying that the model underestimates the proton temperature.
    The statistical details of other CRs are mentioned in Table \ref{tab:statistical_results}.
    \par
   In addition to used value of $\gamma$ (= 5/3, $\gamma_0$), we have also shown the results for $\gamma=1.50$ ($\gamma_1$) and $\gamma=4/3$ ($\gamma_2$) in Figure \ref{fig:plasma_2081}. The profile of all three presented properties showed some changes, with temperature varying the most. With the decrease in $\gamma$ value, the proton temperature increased significantly. $\gamma_1$ and $\gamma_2$ profiles gave a better match for the peak around 60\textdegree\,as compared to $\gamma_0$, but rest of their profile gave high mean error values, specially $\gamma_2$. In the density plot, all three showed similar linear correlation but the mean error of $\gamma_0$ profile were higher at peaks. For magnetic field, all three profiles have equivalent linear correlation and root mean square error. However, $\gamma_0$ profile shows better match than $\gamma_1$ and $\gamma_2$ at the 60\textdegree\, peak. Furthermore, the accuracy of solar wind speed profile also reduced with $\gamma$ value ($rmse$ for: $\gamma_{0}=84.99\, km/s$, $\gamma_{1}=86.12\, km/s$, $\gamma_{2}=87.10\, km/s$). Therefore, slight decrease in $\gamma$ value, from 5/3, might give better match for proton temperature and density but the magnetic field and plasma speed accuracy might decrease.
    
\begin{table*}[t]
\centering
\caption{Statistical results of comparison of model output with the OMNI data at L1. Here, $std\_model$ is standard deviation of model output at 1 AU and $std\_obs$ is standard deviation of the observed OMNI data at L1.}
\label{tab:statistical_results}
\resizebox{\textwidth}{!}{%
\begin{tabular}{l c c c c c c}
\hline
\multicolumn{1}{c}{} & \multicolumn{1}{c}{} & \multicolumn{1}{c}{\textbf{HUX}} & \multicolumn{4}{c}{\textbf{MHD}}    \\ \cline{4-7} 
\multicolumn{1}{c}{\multirow{-2}{*}{\textbf{CR\_MAP}}} & \multicolumn{1}{c}{\multirow{-2}{*}{\textbf{Statistical parameter}}} & \multicolumn{1}{c}{\textbf{Speed (km/s)}} & \multicolumn{1}{c}{\textbf{Speed (km/s)}} & \multicolumn{1}{c}{\textbf{$|$B$|$ (nT)}} & \multicolumn{1}{c}{\textbf{Density (N$_p$cm$^{-3}$)}} & \textbf{Temperature (MK)}                    \\ \hline
 & \multicolumn{1}{l}{}  & \multicolumn{1}{l}{} & \multicolumn{1}{l}{} & \multicolumn{1}{l}{} & \multicolumn{1}{l}{} & \multicolumn{1}{l}{} \\
 & cc & 0.85  & 0.81 & 0.46 & 0.40 & 0.57                    \\
 & rmse  & 74.01 & 80.87 & 4.16 & 20.83 & 0.09               \\
 & std\_model & 123.97 & 135.07 & 4.46 & 19.93 & 0.10        \\
\multirow{-5}{*}{2053\_GONG} & std\_obs  & \multicolumn{2}{c}{119.46} & 2.03  & 4.51 & 0.10   \\ \hline
 & \multicolumn{1}{l}{} & \multicolumn{1}{l}{} & \multicolumn{1}{l}{} & \multicolumn{1}{l}{} & \multicolumn{1}{l}{} & \multicolumn{1}{l}{} \\
 & cc   & 0.64  & 0.73  & 0.20  & 0.39   & 0.40                 \\
 & rmse & 75.21 & 63.58 & 2.57  & 14.74  & 0.07                 \\
 & std\_model   & 80.12 & 71.63 & 1.29   & 12.41 & 0.06         \\
\multirow{-5}{*}{2077\_GONG}    & std\_obs       & \multicolumn{2}{c}{93.33}   & 2.30    & 5.17  & 0.06       \\ \hline
 & \multicolumn{1}{l}{} & \multicolumn{1}{l}{} & \multicolumn{1}{l}{} & \multicolumn{1}{l}{} & \multicolumn{1}{l}{} & \multicolumn{1}{l}{} \\
 & cc & 0.87 & 0.64 & 0.42 & 0.50 & 0.30                        \\
 & rmse  & 41.39  & 84.99  & 2.72  & 7.76  & 0.05              \\
 & std\_model & 69.75 & 100.03 & 2.02 & 8.12 & 0.04             \\
\multirow{-5}{*}{2081\_GONG} & std\_obs & \multicolumn{2}{c}{67.69} & 2.33 & 3.30 & 0.05         \\ \hline
 & \multicolumn{1}{l}{}  & \multicolumn{1}{l}{}  & \multicolumn{1}{l}{}   & \multicolumn{1}{l}{}  & \multicolumn{1}{l}{}    & \multicolumn{1}{l}{} \\
 & cc  & 0.86  & 0.84  & 0.21  & 0.32  & 0.31      \\
 & rmse & 43.58 & 47.81 & 2.58 & 14.33 & 0.06       \\
 & std\_model  & 75.48  & 85.59  & 1.48  & 13.27  & 0.05           \\
\multirow{-5}{*}{2104\_GONG}  & std\_obs   & \multicolumn{2}{c}{82.89}  & 1.96  & 4.37  & 0.06                    \\ \hline
 & \multicolumn{1}{l}{} & \multicolumn{1}{l}{} & \multicolumn{1}{l}{} & \multicolumn{1}{l}{} & \multicolumn{1}{l}{} & \multicolumn{1}{l}{} \\
 & cc  & 0.67 & 0.57 & 0.29 & 0.21 & 0.15          \\
 & rmse & 68.90 & 73.14 & 2.45 & 16.42 & 0.08      \\
 & std\_model & 63.63 & 63.29 & 1.82 & 14.11 & 0.06    \\
\multirow{-5}{*}{2104\_HMI}  & std\_obs & \multicolumn{2}{c}{82.89} & 1.96 & 4.37 & 0.06     \\ \hline
 & \multicolumn{1}{l}{} & \multicolumn{1}{l}{} & \multicolumn{1}{l}{} & \multicolumn{1}{l}{} & \multicolumn{1}{l}{} & \multicolumn{1}{l}{} \\
 & cc & 0.54 & 0.68 & 0.40 & 0.34 & 0.25           \\
 & rmse & 74.35 & 61.86 & 2.85 & 13.44 & 0.06      \\
 & std\_model & 74.37 & 78.63 & 3.08 & 12.25 & 0.04      \\
\multirow{-5}{*}{2202\_GONG} & std\_obs & \multicolumn{2}{c}{73.20} & 1.39 & 3.48 & 0.05 \\
\hline
\end{tabular}%
}
\end{table*}

    \subsection{Solar Wind Interaction Region}
    
    \begin{figure*}
       \centering
       \includegraphics[width = \textwidth]{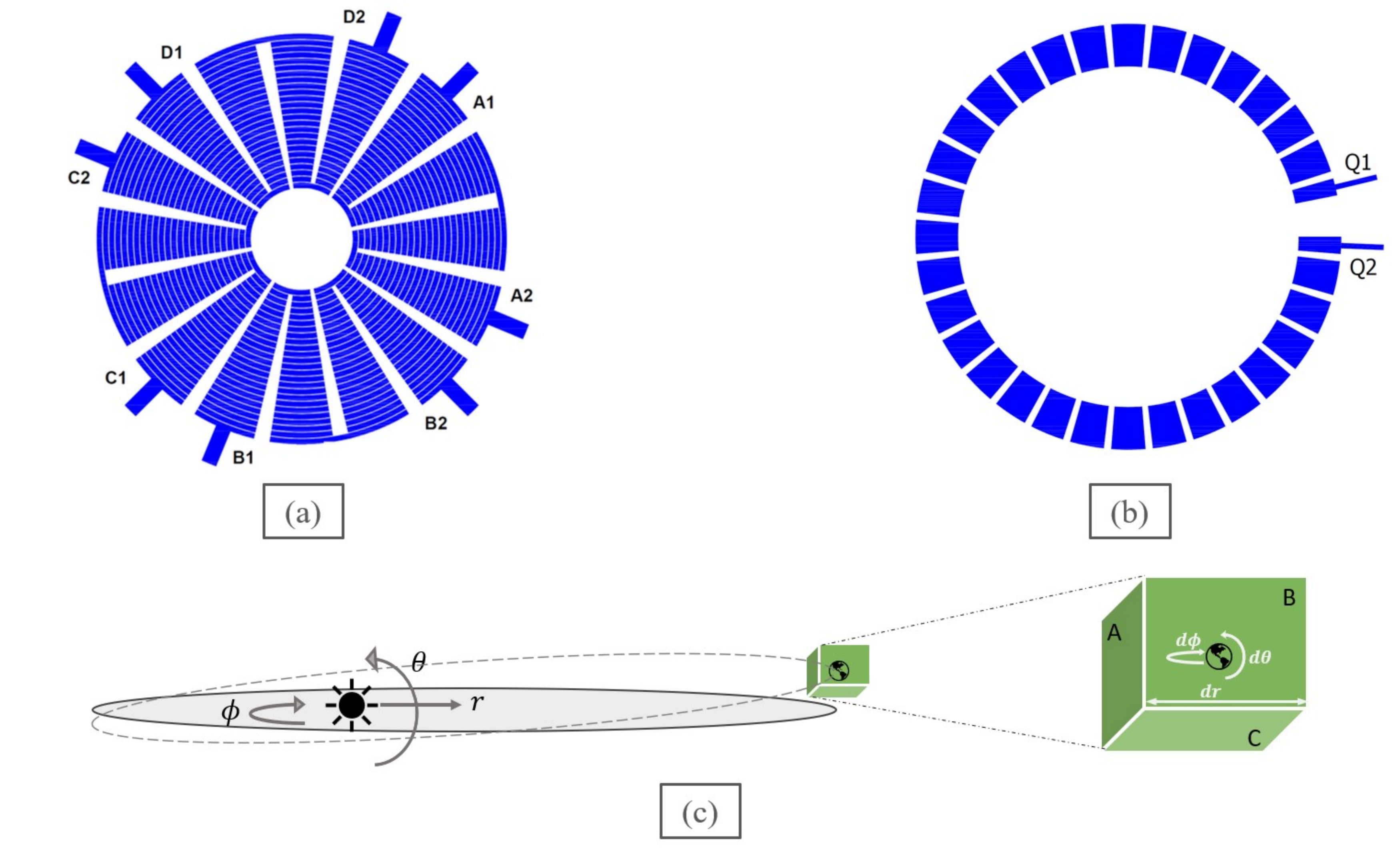}
       \caption{Schematic diagrams of RAE of (a) THA-1, (b) THA-2 of SWIS-ASPEX and (c) the computational surfaces (A, B and C) covering the openings of detectors. Computational surfaces A and B covers the THA-1 and C covers the THA-2. In subplot (c), the solid eclipse represents the equatorial plane in Sun centered frame and the dotted eclipse is the ecliptic plane. The direction of unit vectors in ecliptic plane have been shown and $dr$, $d\theta$ and $d\phi$ are widths of computational surfaces along those directions.}
   \label{fig:swis-aspex}
   \end{figure*}
    
    Apart from determining the plasma properties in the inner heliosphere, SWASTi-SW can also be used to study the high speed solar wind streams (HSSs) and stream interaction regions (SIRs). Additionally, the model can also mimic the observations of in-situ instruments of the upcoming Aditya-L1 mission. For example, Figure \ref{fig:swis-aspex} shows the schematic diagram of Solar Wind Ion Spectrometer (SWIS) which is a subsystem of Aditya Solarwind Particle EXperiment (ASPEX). The three dimensional physics based model allows us to assess the characteristic of SIRs at L1 which can be used as template for directional dependent data acquired by such in-situ payloads (see Section \ref{Section 4} for details).
    \par
    Statistical studies e.g., \citet{Tsurutani2006, Alves2006,Zhang2008} have shown that SIRs/HSSs are chiefly responsible for weak to moderate geomagnetic storms. SIR is produced when HSS interacts with its preceding slower solar wind stream. This interaction causes the formation of compressed plasma density and interplanetary magnetic field (IMF) at the leading edge of the rising section in speed profile.
    \par
    \citet{Belcher1971} classified this interaction into four regions S, S$^{'}$, F$^{'}$ and F i.e., the unperturbed slow wind, accelerating slow wind, decelerating fast wind and ambient fast wind regions, respectively. The compressed S$^{'}$ and F$^{'}$ regions form the SIR with enhanced plasma density and magnetic field magnitude. We observed the same kind of structure formation in predicted plasma properties at L1. Figure \ref{fig:Aspex_SIR} (a) shows the SIR occurring in the first quarter of the CR2081 simulation. Subplots (a1), (a2) and (a3) show the rise in radial velocity, magnetic field magnitude and density in the interaction region. Subplots (a4) and (a5) display the fluctuation in azimuthal component of velocity and flow angle ($\tan^{-1}\bigg(\frac{V_y}{V_x}\bigg)$), where x and y are in HEEQ) during the interaction region. Similar pattern was also predicted for transverse component of velocity vector in \citet[Fig. 17]{Belcher1971}. Furthermore, we observed an additional peculiar trait for plasma flux in the \textbf{meridional and} azimuthal directions which is discussed in detail in Section \ref{Section 4}.

\section{ASPEX: ADITYA L1}\label{Section 4}

The upcoming Indian solar mission, Aditya L1 has seven payloads, four remote sensing and three in-situ instruments. Aditya Solar wind Particle EXperiment (ASPEX) is one of the three in-situ payloads and have multi-directional measurement capabilities. SWIS and SupraThermal \& Energetic Particle Spectrometer (STEPS) are two independent subsystems of ASPEX, which, in turn, have two and six units respectively. The primary scientific objective of SWIS is to study the solar wind plasma particles, whereas that of STEPS is to investigate the suprathermal particles and SEPs \citep{Goyal2018}.

\begin{figure*}
   \centering
   \includegraphics[width = 0.9\textwidth]{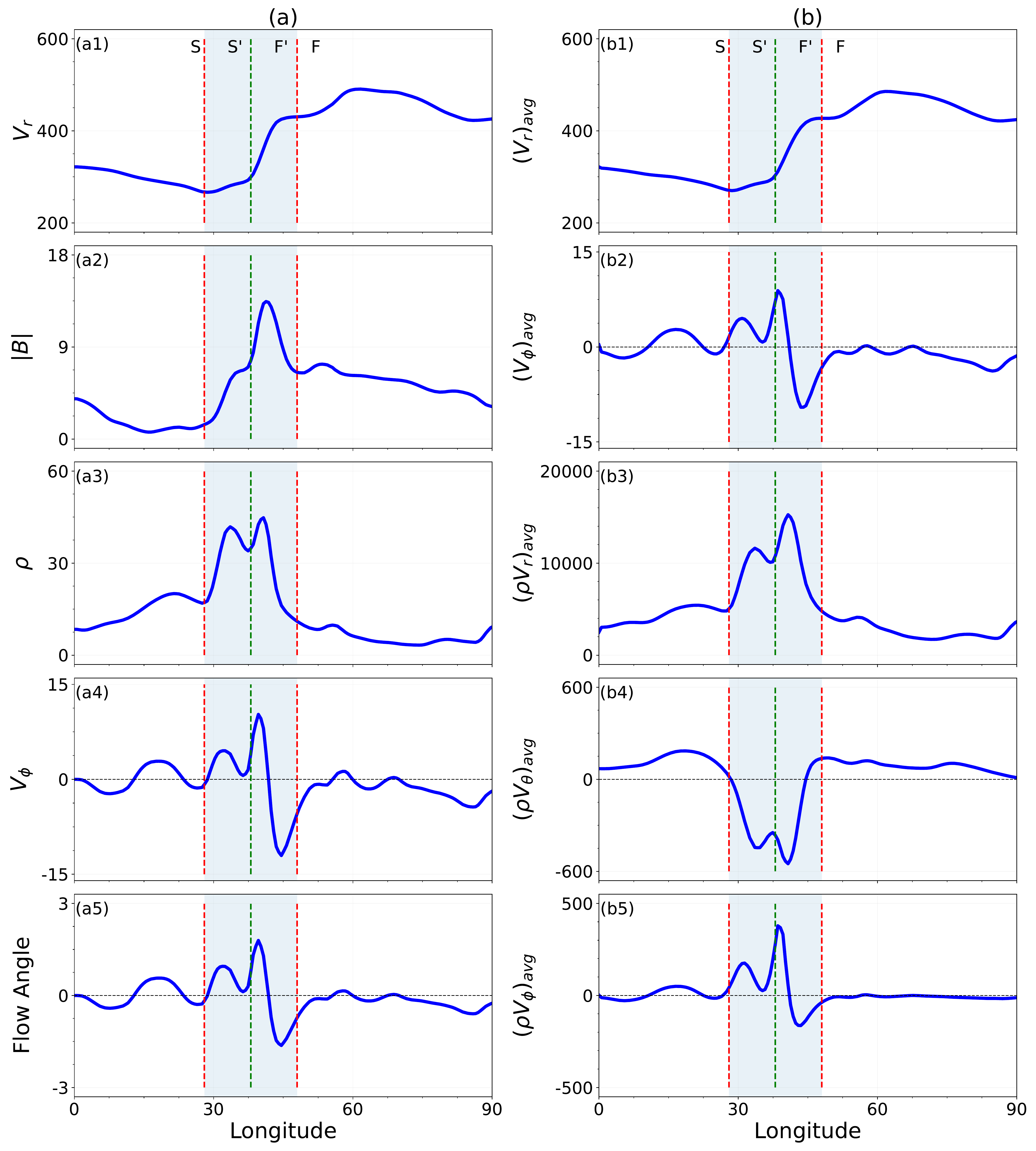}
   \caption{Plots showing the features of plasma properties at L1 corresponding to SIR for CR2081. Plot (a) represents the computed plasma properties at L1 location whereas, plot (b) represents the mass averaged properties at L1. The shaded region (S' and F) shows the SIR where in S' region (between red and green dotted vertical lines) solar wind is getting accelerated and in F region (between green and red dotted vertical lines) plasma is getting decelerated. $V_r$ are $V_\phi$ are plasma velocity in radial and azimuthal direction in km/s. $|B|$ and $\rho$ are IMF magnitude (nT) and proton density ($N_{p}\ cc^{-1}$). Subplot (a5) shows the variation of flow angle in degree. $(V_r)_{avg}$ and $(V_\phi)_{avg}$ are averaged value of $V_r$ and $V_\phi$ (km/s). $(\rho V_{r})_{avg}$, $(\rho V_\theta)_{avg}$ and $(\rho V_\phi)_{avg}$ are averaged proton flux ($N_{p}\ cm^{-3}s^{-1}$) in radial, meridional and azimuthal directions. The X-axis, Longitude, is flipped Carrington longitude same as in Figure \ref{fig:Vel_allCRs} and \ref{fig:plasma_2081}.}
   \label{fig:Aspex_SIR}
\end{figure*}

The two independent units of SWIS, Top Hat 1 (THA-1) and Top Hat 2 (THA-2) will measure the particle flux in energy range of 100 eV to 20 keV. THA-1 have a field of view (FOV) of 360\textdegree$\,$ along the ecliptic plane and will be capable in differentiating the major solar wind ion species (proton and alpha particles). On the other hand, THA-2 will measure total particle flux with 360\textdegree$\,$ FOV in plane perpendicular to ecliptic plane. Each THA has an electrostatic analyzer (ESA) section that selects the incoming ions based on their energies. In THA-1, there is an additional Magnetic Mass Analyzer (MMA) section that deflects the major ions as per their masses to eventually get detected by a Micro Channel Plate (MCP). The position information of the incident charges are derived from the in-house developed Resistive Anode Encoder (RAE). It consists of a number of metallic tracks printed on a PCB material. The position information is derived based on the voltage division across the resistive chains and readout at the end (A1-A2, B1-B2, C1-C2, D1-D2). 
It is to be noted that RAE of THA-1 consists of four quadrants and each resistive chain in a quadrant consists of four sectors that amounts to an angular resolution of 22.5 degrees (16 sectors). This is shown in Figure \ref{fig:swis-aspex} (a).  In addition to the fact that THA-2 is mounted perpendicular to THA-1, there is one more fundamental difference between the THA-1 and THA-2 units. THA-2 does not have any MMA section and hence it is not designed to separate protons and alpha particles. In absence of MMA section, the particles fall on an annular region on RAE that is around the mean radius of the ESA. Therefore, RAE in THA-2 consists of a thin annular strip of a single resistive chain and readouts at the end of the strip. In addition to these differences, THA-2 also has 32 sectors that results in an angular resolution of 11.25 degrees. The schematic of the RAE of THA-2 is shown in Figure \ref{fig:swis-aspex} (b). Therefore, once deployed at L1, SWIS will continuously measure the proton and alpha particles individually in radial and azimuthal directions and integrated flux in the meridional direction.

\par
In this work, we have used our simulation results to synthesize the measurements of SWIS by forming three computational planes (namely A, B and C), each in radial, meridional and azimuthal direction. Each plane is made up of nine grid cells and are at a distance of $dr$ ($\sim$2.8$R_{\odot}$), $d\theta$ (1\textdegree) and $d\phi$ (1\textdegree) from the L1 grid cell, in their respective unit vector directions. Figure \ref{fig:swis-aspex} (c) displays the computational planes that cover the FOV of SWIS and the L1 grid cell is represented by the Earth's logo. The values at nine grid cells are used to find the averaged value of quantities at L1 which are plotted in Figure \ref{fig:Aspex_SIR} (b).
\par
We used the above mentioned averaging technique to imitate the observation of SWIS with the aim to show how its multi-directional measurement capability could lead to more accurate detection of SIRs. For demonstration purpose, we chose a well-defined SIR in the first quarter of CR2081. Figure \ref{fig:Aspex_SIR} subplots (b1) and (b2) are of averaged radial and azimuthal components of velocity and they are very much similar to their value at the L1 grid cell, implying that the employed technique is satisfactory. Figure \ref{fig:Aspex_SIR} (b3), (b4) and (b5) are of proton flux density in radial, meridional and azimuthal directions, respectively. In all the three flux plots, there is significant change in S$^{'}$ and F$^{'}$ (shaded) regions. The radial flow has increased, meridional flow has changed its direction and azimuthal flow is fluctuating rapidly in the shaded S$^{'}$ and F$^{'}$ regions. Similar features were also observed for interaction regions at 120\textdegree$\,$ and 300\textdegree$\,$ of CR2053, 60\textdegree$\,$ of CR2077 and 240\textdegree$\,$ of CR2104.
This collective feature occurring in proton flux profile in three directions, specially the fluctuations in azimuthal flow and meridional flow, can be called as the characteristic feature of SIRs. The observation of such multi-directional features can be used to detect SIR events at L1, along with rise in density and magnetic field. This will result in a more reliable and precise detection of SIRs at L1.
\par
To observe the above mentioned composite features in plasma flux due to SIR, the spacecraft should have the facility to collect the data continuously in all the three directions. ASPEX has that directional capability and can detect the rise in plasma density, plasma flux in radial as well as fluctuation of flux in azimuthal direction. Along with the on-board magnetometer, Aditya L1 will have the ability to measure the rise in plasma density, IMF magnitude and fluctuations in longitudinal direction. The combined observation of the three specified features will provide a better SIR detection functionality. 


\section{Summary and Discussion}\label{Section 5}

In this work, we presented an indigenous physics-based solar wind forecasting model for inner heliosphere which uses an adapted semi-empirical approach for initial boundary condition. This 3D model has been developed with an intention to run on a personal workstation in reasonable computational time with adequate accuracy. On an average, the coronal domain takes around 2 hours and inner-heliospheric domain takes around 6.5 hours for one complete CR run on a 48 core processor. This computational time is for the resolution and setup mentioned in Section \ref{Section 2}. To reduce this computational time, user can opt for a lower angular resolution of 2\textdegree$\,$. However, due to lack of sampling in the longitudinal and latitudinal directions, features of SIR (as depicted in Figure \ref{fig:Aspex_SIR}) would be rather diffused and their identification would be troublesome. Therefore, when SIR assessment is not the objective, then the user can opt for a lower angular resolution.
\par
In this study, two types of magnetograms have been used to validate the results with observation, GONG and HMI. Being the only observational input, accuracy of magnetogram directly effects the veracity of model prediction. Both GONG and HMI measures LOS component and provides radial component of magnetic field on the solar surface. They do so by dividing the observed magnitude by the cosine of the angle from the disk center and assuming that the fields are purely radial at photosphere. But this approximation is sensible only in those regions where the field is not strong enough to resist the fluid forces, like for quiet and weak active regions and not for strong active regions. Therefore, an inevitable discrepancy is always present in the model's input which ultimately gets reflected in final results.
\par
In SWASTi-SW, the initial boundary condition of MHD domain is based on an adapted version of original WSA relation \citep{Arge2003}. We generalized the empirical relation of plasma speed by successfully reducing one independent parameter. This generic approach makes the WSA model independent of the choice of the grid resolution in inner-heliospheric domain, specially in latitudinal and longitudinal direction.
\par
An effort to optimize the value of one other independent parameter using HUX algorithm was also demonstrated. Though this method uses observational data and can't be directly used for forecasting purposes, a long term study using this approach would be helpful in finding the optimal set of values of independent parameters. A possible future work would be to use such extensive data set from long term studies to automatize the choice of free parameters for any given Carrington rotation period. 
\par
The results from comparison suggests that SWASTi-SW is capable in forecasting the ambient solar wind properties at L1, specially the correlation of plasma speed is high (upto $cc=0.84$). Model does overestimate (underestimate) the value of plasma density (temperature) and this is something that should be explored in future works. We showcased that a reduced value of specific heat ratio ($\gamma$) leads to additional heating, but it also decreases the density and magnetic field intensity. Therefore, a spatially varying relation of $\gamma$ can be a reasonable alternative for better match. Another rational approach could be the introduction of anisotropic pressure terms in MHD equation to incorporate more insightful physics.
\par
We demonstrated the directional dependent features (e.g., velocity, proton flux, flow angle, etc.) of SIRs using our model and also presented a synthetic measurement to mimic the observations of the SWIS. At this point, we again reiterate that SWIS-THA1 does not have very good latitudinal coverage but exceptional azimuthal coverage. On the other hand, SWIS-THA2 has exceptional latitudinal coverage but limited azimuthal coverage. Therefore, by combining THA-1 and THA-2 measurements, one can pick up the signatures of SIRs in three dimensions. Additionally, \citet{Rout2017} showed that for geoeffectiveness, CIR azimuthal flow angle is mostly within 6 degree. The advantage here is, ASPEX can capture the signatures of the arrival of SIRs/CIRs at the L1 point at all azimuthal and elevation angles - some of these will turn out to be geoeffective and some will not. In addition, the STEPS subsystem that covers six directions, can detect the energetic particles arriving from the SIR/CIR shock fronts. Therefore,  SWIS and STEPS together along with the modeling outputs can be very important for SIR/CIR investigations and can pave the way for the forecasting of their arrival. This framework will therefore complement the upcoming ISRO mission, Aditya-L1. Specifically SWASTi-SW will compliment the in-situ measurements of solar wind properties by ASPEX and MAG.



\begin{acknowledgments}
We thank the support provided by IIT Indore to carry out this work. PM would thank the financial support provided by the Prime Minister's Research Fellowship. BV and DC would like to kindly acknowledge the support from the ISRO RESPOND grant number: ISRO/RES/2/436/21-22. The work of DC is supported by the Department of Space, Government of India. We would also like to place on record the untiring efforts and contributions of the whole ASPEX team at PRL in the realization of the payload. The support and inputs from the Space Application Center (SAC), Ahmedabad,  various ISRO Centers, splinter groups, review committees, Aditya Science Working Group are duly acknowledged. The guidance and support from the Director, PRL towards the ASPEX project are invaluable.
\par
The used GONG and HMI synoptic magnetograms maps can be freely obtained from https://gong.nso.edu/data/magmap/crmap.html and http://jsoc.stanford.edu/HMI/Magnetograms.html, respectively. The OMNI data are taken from the Goddard Space Flight Center, accessible at https://spdf.gsfc.nasa.gov/pub/data/omni/. {\sc pfsspy} python package used in this work for PFSS modelling is available at https://pfsspy.readthedocs.io/en/stable/. The {\sc pluto} code used for MHD simulation can be downloaded free of charge from http://plutocode.ph.unito.it/.
\end{acknowledgments}

\bibliography{Bibliography}{}
\bibliographystyle{aasjournal}



\end{document}